# Direct observation of cation-dependent polarisation switching dynamics in fluorite ferroelectrics

Kousuke Ooe[1,2*], Yufan Shen[3], Kazuki Shitara[2], Shunsuke Kobayashi[2], Yuichi Shimakawa[3], Daisuke Kan[3†], Joanne Etheridge[1,4]

[1]School of Physics and Astronomy, Monash University, Clayton, VIC 3800, Australia.
[2]Nanostructures Research Laboratory, Japan Fine Ceramics Center, Nagoya, Aichi 456-8587, Japan.
[3]Institute for Chemical Research, Kyoto University, Uji, Kyoto 611-0011, Japan.
[4]Monash Centre for Electron Microscopy, Monash University, Clayton, VIC 3800, Australia.
* Corresponding author. Email: kousuke.ooe@monash.edu (K.O.)
† Present address: Division of Applied Chemistry, Graduate School of Engineering, The University of Osaka, Suita, Osaka 565-0871, Japan.

## Abstract

Fluorite ferroelectrics are exciting candidates for next-generation non-volatile memory devices because their unique ferroelectric mechanism, which arises from unconventional oxygen displacements, permits ferroelectricity with minimal thickness constraints. However, the polarisation switching mechanism remains the subject of intense debate due to a limited understanding of the atomic-scale dynamics which are extremely challenging to detect and measure. Here, we observe directly the polarisation switching pathways by visualising oxygen site dynamics in $ZrO_2$ and $Hf_{0.5}Zr_{0.5}O_2$ freestanding membranes using an advanced atomic-column imaging technique—optimum bright-field scanning transmission electron microscopy. We observe that the 180° and 90° polarisation pathways involve different nonpolar intermediate states with distinct spatial scales. Coupled with density functional theory, we also reveal how different cation species in fluorite oxides impact the accessible polarisation switching pathways. Our atomic-level insights into the polarisation switching dynamics open new avenues for the advanced engineering of fluorite ferroelectric materials and resulting memory devices.

# Introduction

Ferroelectric materials, characterised by a spontaneous polarisation that can be reversed by an external electric field, have long been a focus of research in materials science [1]. Their unique ability to switch polarisation provides a wide range of applications, including non-volatile memory devices [2]. Traditionally, perovskite oxides such as $BaTiO_3$ and $Pb(Zr,Ti)O_3$ have been known as typical ferroelectric materials [3]. However, these materials face intrinsic limitations when scaled to nanometre dimensions; for instance, the ferroelectric properties in perovskites generally degrades at film thicknesses of tens of nanometres, which hampers further miniaturisation of memory devices [4].

A breakthrough in the field came with the discovery of ferroelectricity in fluorite-type oxides such as $HfO_2$, $ZrO_2$, and $Hf_{0.5}Zr_{0.5}O_2$ (HZO) [5]. Unlike conventional perovskites, ferroelectricity in these materials arises from shifts of oxygen anions rather than from the displacement of cations within a crystalline structure. $HfO_2$, already a standard high-$\kappa$ material in complementary metal-oxide-semiconductor (CMOS) processes [6], demonstrated robust ferroelectric switching in Si-doped films, as first reported by Böscke et al. [7]. This discovery not only spurred extensive investigations into the fundamental mechanisms governing ferroelectricity in these systems but also highlighted their potential for enabling devices with ultrathin active layers—down to nanometre or even just a few atomic layers—thus overcoming the critical thickness limitations inherent to perovskite ferroelectrics [8].

In fluorite-type oxides, the ferroelectric phase is typically associated with a non-centrosymmetric orthorhombic structure (space group $Pca2_1$), in which displacements of oxygen ions with respect to the heavy cations (e.g., Hf and Zr) produce a net dipole moment [7]. Notably, these materials often exhibit polycrystalline structures, i.e., a texture of nanometre-sized grains with different crystalline phases, including metastable polar and nonpolar phases due to the near-degeneracy of these phases [9,10]. Consequently, unveiling these crystalline structure characteristics and their transitions, and engineering polarisation switching in the ferroelectric phase of fluorite oxides is critical not only for advancing fundamental understanding but also for optimising their performance in memory applications because these dynamics are inevitably related to the wake-up and/or fatigue phenomena [8].

Scanning transmission electron microscopy (STEM) has been emerging as a powerful tool in this regard. Modern aberration-corrected STEM techniques enable direct imaging of individual atomic columns with sub-Ångström resolution, thereby providing detailed insights into the local atomic structure [11]. High-angle annular dark-field (HAADF) STEM is a common STEM imaging technique for materials characterisation that excels in visualising heavy elements such as cations, and hence is useful for detecting the cation-induced polarisation structures [12]. As for the oxygen sites, which plays a pivotal role for the fluorite-type ferroelectrics, phase-contrast-based techniques can be applied to visualise them. By means of these, recent STEM analysis of fluorite ferroelectric materials not only identified crystalline phase of individual nano-sized grains [13] but also observed polarisation structures, some of which visualised the structure evolution such as polar-nonpolar phase transition and polarisation switching triggered by an electron irradiation or electric-field biasing [14–16]. However, the detailed atomic structures and dynamics of these transition pathways, including intermediate states, remain unclear, and consequently the means to selectively facilitate or suppress specific pathways to obtain the desired ferroelectric phase and polarisation switching paths remain unknown.

In this study, we unveil the atomistic pathways of 180° and 90° polarisation switching and their intermediate structures involved in the fluorite ferroelectrics. Critically, we reveal the accessible pathways to be cation-species-dependent, by comparing the atomic structure dynamics of freestanding $ZrO_2$ and HZO membranes. To investigate the detailed polarisation structures and switching pathways, we employ an optimum bright-field (OBF) STEM technique [17], which can visualise both oxygen and cation sites simultaneously with high sensitivity. Our approach first demonstrates that OBF STEM can effectively visualise the local atomic structure within these membranes, resolving both the polar and nonpolar phases. Building on this static analysis, we then explore the dynamic behaviour of these atomic structures under high beam-current electron irradiation. In particular, our observations in $ZrO_2$ membranes reveal the 180° and 90° polarization switching within the ferroelectric structure is mediated by reversible phase transitions between polar and nonpolar states. Furthermore, we perform a higher-speed OBF STEM data acquisition, which allows us to capture two different nonpolar intermediate structures along the 180° and 90° polarisation switching pathways, providing unprecedented insight into the transient states that occur during the switching process. We also perform the observation of an HZO membrane, shedding light on how the different atomic species of cation sites impacts the dynamic behaviour regarding the phase transition and polarisation switching. The energetic landscape governing these transitions and switching processes is elucidated by the density functional theory (DFT) calculations, complementing our experimental findings. In addition to the detailed structure evolution analysis during the phase transition and polarisation switching of $ZrO_2$ and HZO, our approach offers a critical guideline to engineer these dynamics and realise the advanced fluorite ferroelectrics.

# Results and Discussion

**Visualisation of the static polarisation structure at the atomic scale**

The projected atomic structure models along the [001] zone-axis of nonpolar tetragonal (space group *P*4$_2$/*mmc*), polar orthorhombic (space group *Pca*2$_1$), and anti-polar orthorhombic (space group *Pbca*) phases in the fluorite ZrO$_2$ and HZO are shown in Figure 1a. In the tetragonal phase, cation (Hf/Zr) sites form a cubic-like lattice along this projection and oxygen sites are located in the centre of this cation lattice, giving a nonpolar structure as shown in Fig. 1a. In the orthorhombic phase, on the other hand, the distance in the [100] direction between the tetragonal (100) and (200) cation planes alternates from narrow (2.36 Å) to wide (2.69 Å), in this projection. Moreover, in the wide layer, the oxygen site shifts off-centre within the wide cation lattice in the [010] direction, while the oxygen sites in the narrow layer remain centred. This polarisation due to oxygen shifts causes the ferroelectricity in (H)ZO (where we use (H)ZO as shorthand to refer to both ZrO$_2$ and HZO). These fluorite type oxides have distinct polarised atomic layers in between oxygen-centred nonpolar spacers, making the polar domain stable even down to a single atomic layer in width [18]. In the anti-polar orthorhombic phase, the direction of the oxygen shift alternates as shown in Fig. 1a, resulting in the anti-polar behaviour.

We first analysed these features inside the (H)ZO membranes using OBF STEM. The OBF STEM technique can visualise atomic structures including both heavy and light elements with high sensitivity, making it powerful for probing the oxygen-induced polarisation structures described above. For the structure analysis of ZrO$_2$ and HZO, nanometre-thickness membranes were fabricated using the pulsed laser deposition (PLD) method and then transferred onto TEM specimen grids, with details described in the Method section. Figs. 1b-d show the resulting experimental OBF images of each phase inside the ZrO$_2$ (5 and 10 nm thicknesses) and HZO (5 nm thickness) membranes, visualising both the Hf/Zr and O atomic sites. As shown in Fig. 1b, in the nonpolar tetragonal phase observed in the 5 nm ZrO$_2$ membrane, the higher intensity spots, which correspond to the Hf/Zr sites, are arranged in a square lattice in projection, and lower intensity spots, corresponding to the oxygen sites, are located in the centre of this square cation lattice. On the other hand, in the orthorhombic phases, the oxygen sites show a dumbbell-like elongated intensity distribution compared with the tetragonal phase and are located in the off-centred position inside the wider-spaced layers of the cation lattice as shown in Figs. 1c, d. It is also evident that the oxygen shift is aligned along one direction in the polar orthorhombic phase (Fig. 1c), which was observed in the 5 nm thick HZO membrane, whereas the oxygen shift direction alternates 180-degree in the anti-polar phase observed in the 10 nm ZrO$_2$ membrane (Fig. 1d). These result shows that the OBF STEM technique is promising for probing the atomic-scale oxygen-triggered polarisation structures in the fluorite oxides.

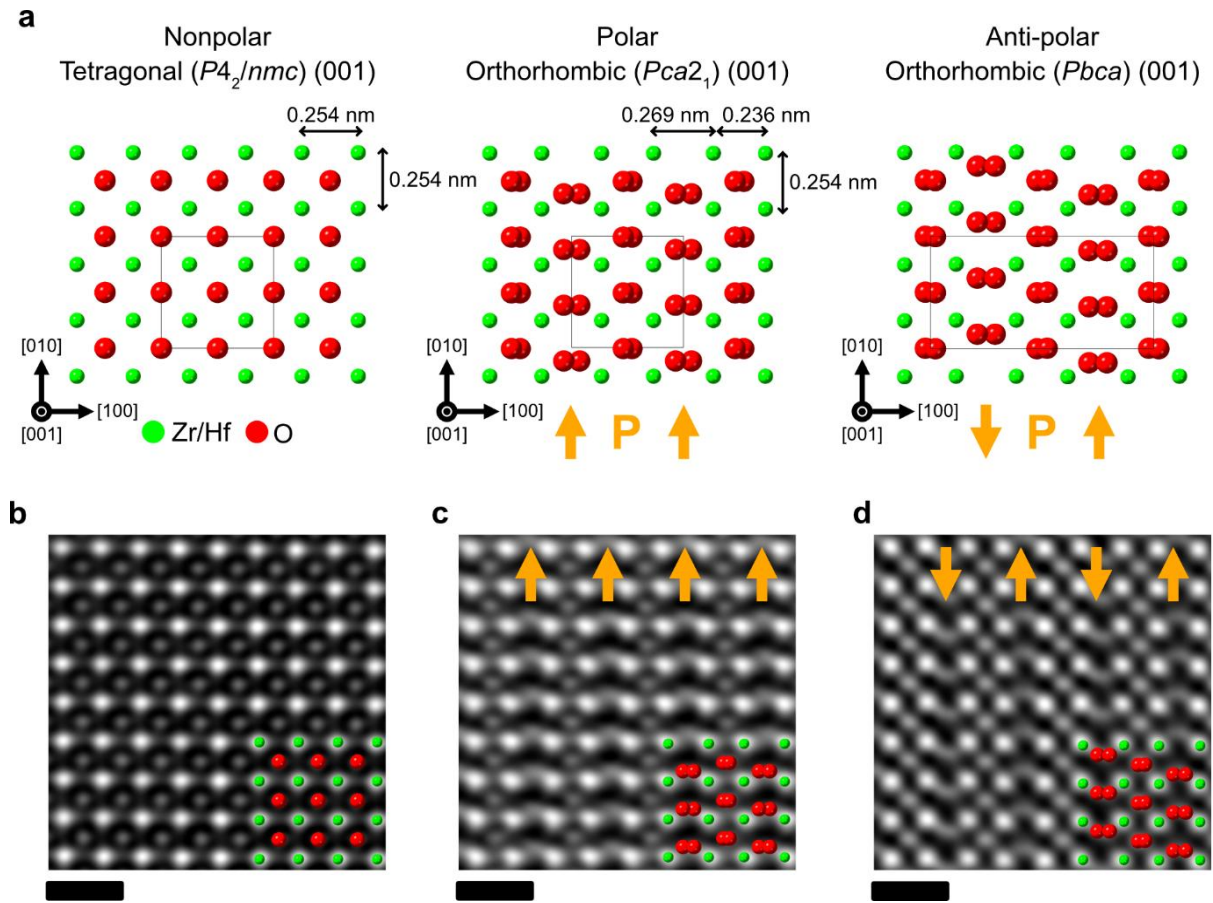

**Figure 1 Atomic structure model and atomic-resolution OBF STEM images of polar/non-polar phases in (H)ZO membranes.** (a) Atomic structure models of nonpolar tetragonal, polar orthorhombic, and anti-polar orthorhombic phases along the projection perpendicular to the (001) plane. Cation (Zr/Hf) and oxygen sites are shown as green and red spheres, respectively. The projected cation lattice distances are also shown, using values obtained from DFT calculations. Experimental OBF STEM images of (b) tetragonal in a ZrO$_2$ membrane (5 nm thick), (c) polar orthorhombic in a HZO membrane (5 nm thick), and (d) anti-polar orthorhombic phases in a ZrO$_2$ membrane (10 nm thick). Yellow arrows indicate the polarisation direction of each atomic layer in the orthorhombic phases. The atomic structure models are also overlayed. Scale bar: 1 nm.

# Reversible phase transition and polarisation switching under high electron dose illumination

In electron microscopy, electron illumination can cause dynamic structure changes inside materials such as phase transitions [19,20], grain boundary migration [21], and polarisation switching [22]. In the case of fluorite ferroelectric materials such as (H)ZO, it is reported that electron irradiation can trigger the phase transition and/or polarisation switching [14]. Observing these structure evolutions and their transition pathway at the atomic scale is critical for understanding the ferroelectric mechanism and how it might be manipulated and optimised for practical applications. Here, we use OBF STEM with a high illumination beam current to capture the dynamic behaviour of our ferroelectric membrane sample at the atomic-scale. In typical STEM measurements, such as shown in Fig. 1, the electron beam current is around 10-20 pA. Here, we deliberately use a 100 pA probe current to stimulate the phase transition and record as a movie, a sequence of images from the same field-of-view (FOV).

To establish a baseline, we first use a relatively low probe current of 15pA. Fig. 2a and b show the corresponding OBF STEM image of the nonpolar tetragonal grain of the 5 nm thickness $ZrO_2$ membrane, as discussed in Fig. 1. We obtained thirty images sequentially with a scanning speed of 500 ns/pixel with a 2048 × 2048 pixels scan (equivalent to three seconds per frame or 0.33 frames per second) from the same field-of-view, resulting in $2.7 \times 10^3$ e$^-$/Å$^2$/frame as a total dose per area. We then averaged these images after aligning them to correct for stage drift. During the sequential acquisition, the tetragonal grain retained its original atomic arrangement as shown in Supplementary Movie S1, indicating that the electron irradiation at 15pA did not change the atomic structure.

We then increased the probe current to a high current of 100 pA by changing monochromator focus value only (while keeping all other electron-optical parameters and resultant observation condition the same) and obtained another image sequence from the same initially tetragonal grain observed above, with results shown in Figs. 2c-e. Here, the electron dose per area was also increased to $1.8 \times 10^4$ e$^-$/Å$^2$/frame. At 69 seconds, the tetragonal grain changed into the orthorhombic phase that has polarised oxygen shifting off centre relative to the cation lattice along the tetragonal [010] direction, which is evident in the OBF STEM image (see also Supplementary Movie 2). After an extra scan, this structure changed back to the original tetragonal structure reversibly.

After several scans taking a further ~30 seconds, the tetragonal grain changed back to an orthorhombic phase (at 102 seconds), however, this time the polarisation structure is still along tetragonal [010] direction but has different polarisation ordering from what was observed in Fig. 2c. After a further beam exposure (at 105 seconds), this grain remained in the orthorhombic phase but a 180-degree polarisation switching occurred in the two leftmost atomic layers shown in Fig. 2d. Before and after this polarisation switching, the arrangement of the cation lattice remained unchanged, but the oxygen sites are shifted locally relative to the Zr cation lattice. After an additional scan, this orthorhombic phase changed back to the tetragonal phase reversibly, with the oxygen located at the centre of the cation lattice. As shown in the Supplementary Movie S3, the 180-degree polarisation switching and phase transition between tetragonal and orthorhombic phases were observed multiple times and were reversed in the subsequent scans.

After additional electron exposure, 90-degree polarisation switching also started to be triggered as shown in Fig. 2e and Supplementary Movie S4. The grain had a tetragonal atomic structure at 216 seconds and then, after a single scan, became a polarised orthorhombic structure where the oxygen sites are shifted in the tetragonal [100] direction with reference to the cation sites (horizontal direction in the figure orientation). At 222 seconds, the orthorhombic structure has changed into an orthorhombic structure polarised along tetragonal [010] direction, showing a different type of switching path from previous results. As discussed in Fig. 1, the polar orthorhombic structure has a uniform cation-cation distance along the polarisation direction and alternately wide and narrow distances perpendicular to it. During this 90-degree switching process, not only the oxygen sites but also the cation sites change their positions, requiring a larger shift of the atomic positions than that required in the 180-degree switching. The distinct change of cation lattice spacing is also confirmed by the cation interatomic distance measurement based on the simultaneously obtained ADF images as shown in Supplementary Figure S3. Here it is clearly visualised that the orthorhombic phase has two different cation interatomic distances alternately perpendicularly to the polarisation direction while the tetragonal phase has uniform interatomic distances. Furthermore, at some scan frames, the grain contains both tetragonal and orthorhombic phases simultaneously, which is confirmed via the oxygen position in OBF STEM images and cation interatomic distances in ADF STEM images as shown in Supplementary FigureS4.

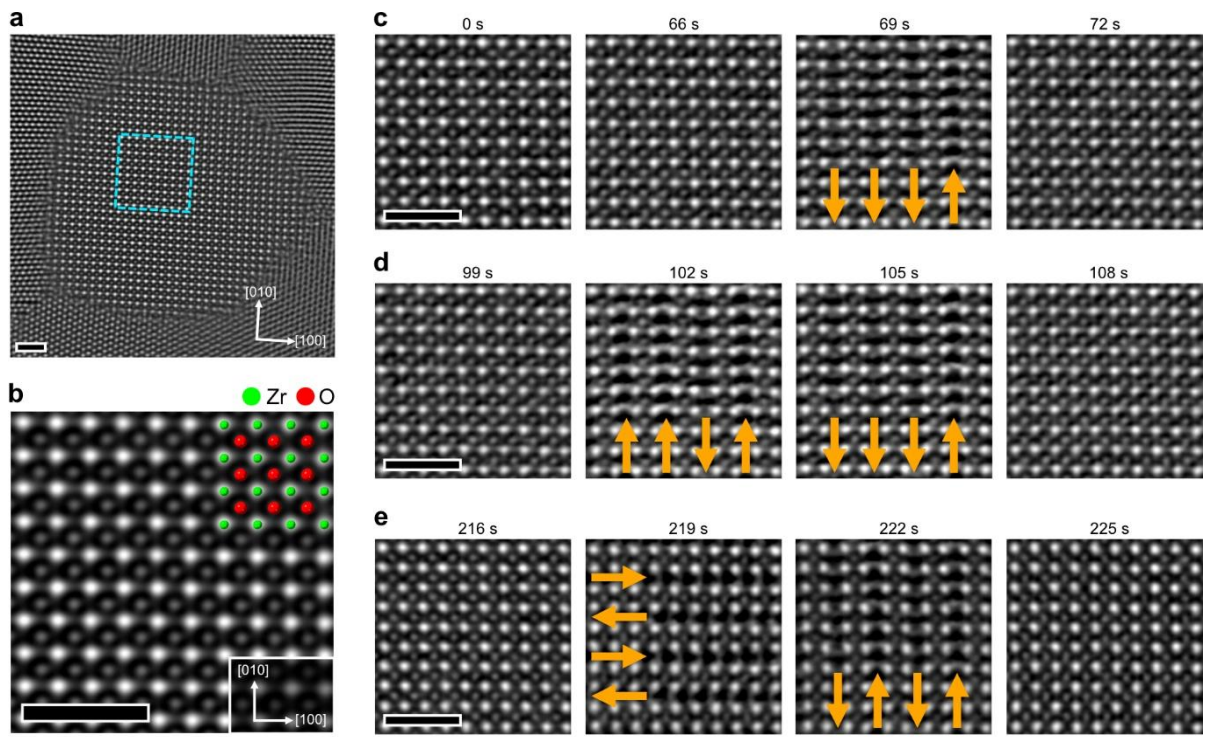

**Figure 2 Direct observation of reversible phase transition and polarisation switching in ZrO$_2$.** (a) OBF STEM image of a tetragonal grain in the ZrO$_2$ membrane. Scale bar: 1 nm. (b) OBF STEM image obtained by cropping and magnifying the cyan-dotted rectangular region shown in (a). The atomic structure model is overlaid. Scale bar: 1 nm. Sequentially obtained OBF STEM images which show (c) (0-72 secs) the phase transition between tetragonal and orthorhombic phases, (d) (99-108 secs) 180-degree polarisation switching, and (e) (216- 225 secs) 90-degree polarisation switching. The time count is set so that the first image acquisition is 0 second just after increasing the probe current to 100pA. Scale bar: 1 nm.

**Intermediate structures during the polarisation switching**

In the previous section, we observed reversible phase transitions and polarisation switching in the ZrO$_2$ membrane by means of OBF STEM with a high probe current. To provide a detailed understanding of the polarisation switching pathway, we performed OBF STEM measurements with a higher-speed scan. With a scan speed of 200 ns/pixel and sampling of 1024×1024 pixels, we obtained sequential OBF images at 0.0629 seconds/frame or 16 frames-per-second. Fig. 3a shows the tetragonal grain observed under an initial 15 pA probe current illumination (5.5 × 10$^2$ e$^-$/Å$^2$/frame), which shows the nonpolar oxygen-centred atomic configuration consistent with the result shown in Fig. 2a. We focus on the single atomic layer along the tetragonal [100] direction as highlighted by the cyan dotted rectangular in Fig. 3a.

Following the initial low current OBF-STEM image of the tetragonal grain, we increased the illumination current to 150 pA (5.5 × 10$^3$ e$^-$/Å$^2$/frame) and took a higher-speed time-series OBF-STEM images. Fig. 3b and Supplementary Movie S5 show the sequence of OBF STEM images, focusing on the single atomic layer inside the grain. As observed in the lower probe current (15pA) condition, the layer shows the tetragonal structure at 0.0 second. At 0.1887 seconds, while the atomic layer mostly has the tetragonal structure, it can be seen that some of the oxygen sites have started to shift away from the centre of the cation lattice, which we denote as the T' phase. At 0.3774 seconds, the oxygen sites have shifted completely in one direction parallel to tetragonal [100] and the atomic layer has changed to the polarised orthorhombic structure as observed in previous sections. After five scans (at 0.9435 seconds), the oxygen sites on the right-hand side of the layer started to shift in another direction along tetragonal [-100] whereas the left-hand side of the layer still shows the polarisation toward the [100] direction. After a further scan (at 1.0064 seconds) the left polarised domain propagated into the left-hand direction. Here, the area between the right- (O(→)) and left- (O(←)) polarised domains has a transient atomic structure (O') during the 180-degree polarisation switching, where the image intensity at the oxygen site is lower than that of the polar orthorhombic phase and distributed either side of the vertical tetragonal (100) and (200) planes of the cation lattice.

During the 180-degree domain switching in the fluorite oxide materials such as (H)ZO, it has been suggested that the domain wall has a nonpolar interlayer between two oppositely polarised domains, which has orthorhombic *Pbcm* structure for the head-to-head boundary and tetragonal *P4$_2$/nmc* structure for the tail-to-tail boundary [23]. To investigate this, we compare our experimental OBF STEM images with simulated images of nonpolar tetragonal, polar orthorhombic (polarised to right and left directions respectively), and non-polar orthorhombic phase (space group *Pbcm*) within single layers, where the polarised oxygen shift occurs when the structure has polar phase, as shown in Fig 3c. In contrast to the tetragonal and polar orthorhombic phases, in the nonpolar orthorhombic *Pbcm* structure, the oxygen site splits into two sites with a half occupancy (shown as half-red and half-white spheres in the structure model) [24], appearing as even lower intensity spots in the simulated OBF image. The experimental OBF image at 1.0064 seconds shown in Fig. 3b corresponds approximately to the simulated intensity distribution for the nonpolar orthorhombic phase, which was suggested to appear on the head-to-head domain boundary from the DFT calculations discussed in the literature [23]. After another scan (at 1.0693s) the left-polarised domain extends into the left-hand side of the layer and a few oxygen sites at the left edge of FOV show diffuse intensity, which is denoted as T' phase because it appears intermediate to the tetragonal phase. At 1.6983 seconds this atomic layer turns back completely to the tetragonal phase, where the oxygen sites are located in the centre of the cation lattice. This result shows that the oxygen site is bridging across the cation lattice plane for the 180-degree switching via the nonpolar orthorhombic *Pbcm* phase.

We next show higher-speed OBF STEM imaging of 90-degree polarisation switching. Fig. 3d shows another tetragonal grain in the same $ZrO_2$ membrane, whose atomic structure was identified by low probe current OBF STEM imaging. This grain first showed 180-degree domain switching after increasing the probe current and then started 90-degree switching subsequently as also shown in Fig. 2e. Fig. 3e shows the time-series of 90-degree switching, starting from the vertically polarised structure, whose time is counted as the 0.0 second frame (see also Supplementary Movie S6). In this image, the oxygen sites are shifted along the tetragonal [010] direction (vertically with respect to the cation sites). In the next scan (0.0629 seconds), these oxygen sites are shifted to the centre of cation lattice, showing the tetragonal phase. At 0.1258 seconds, the oxygen site within a single layer has shifted to the [100] direction (right-hand side), showing the locally left-polarised orthorhombic phase within that monolayer. After another scan at 0.1887 seconds, oxygen sites within the same layer shifted back to the centre of the cation lattice, corresponding to the tetragonal phase again. We illustrate this sequential structure evolution as a series of atomic structure models below each OBF STEM image, corresponding to the region in the cyan dotted box as highlighted in Fig. 3e. This result shows that the oxygen sites are moving around inside the cation sublattice during the 90-degree polarisation switching: the shifted oxygen sites within the polar orthorhombic phase move to the centre, and then shift perpendicularly to the other side, indicating that the 90-degree switching occurs via the nonpolar oxygen-centred tetragonal phase.

From these results, our OBF STEM observations show that the oxygen site behaves differently between the 180- and 90-degree domain switching via distinct nonpolar intermediate states, as shown schematically in Fig. 3f. In the case of 180-degree switching, several oxygen sites collectively cross the cation lattice plane and form the nonpolar intermediate phase which extends across a few nanometres. In contrast, in the case of the 90-degree switching, each oxygen site is confined and moves locally within the individual cation sublattice.

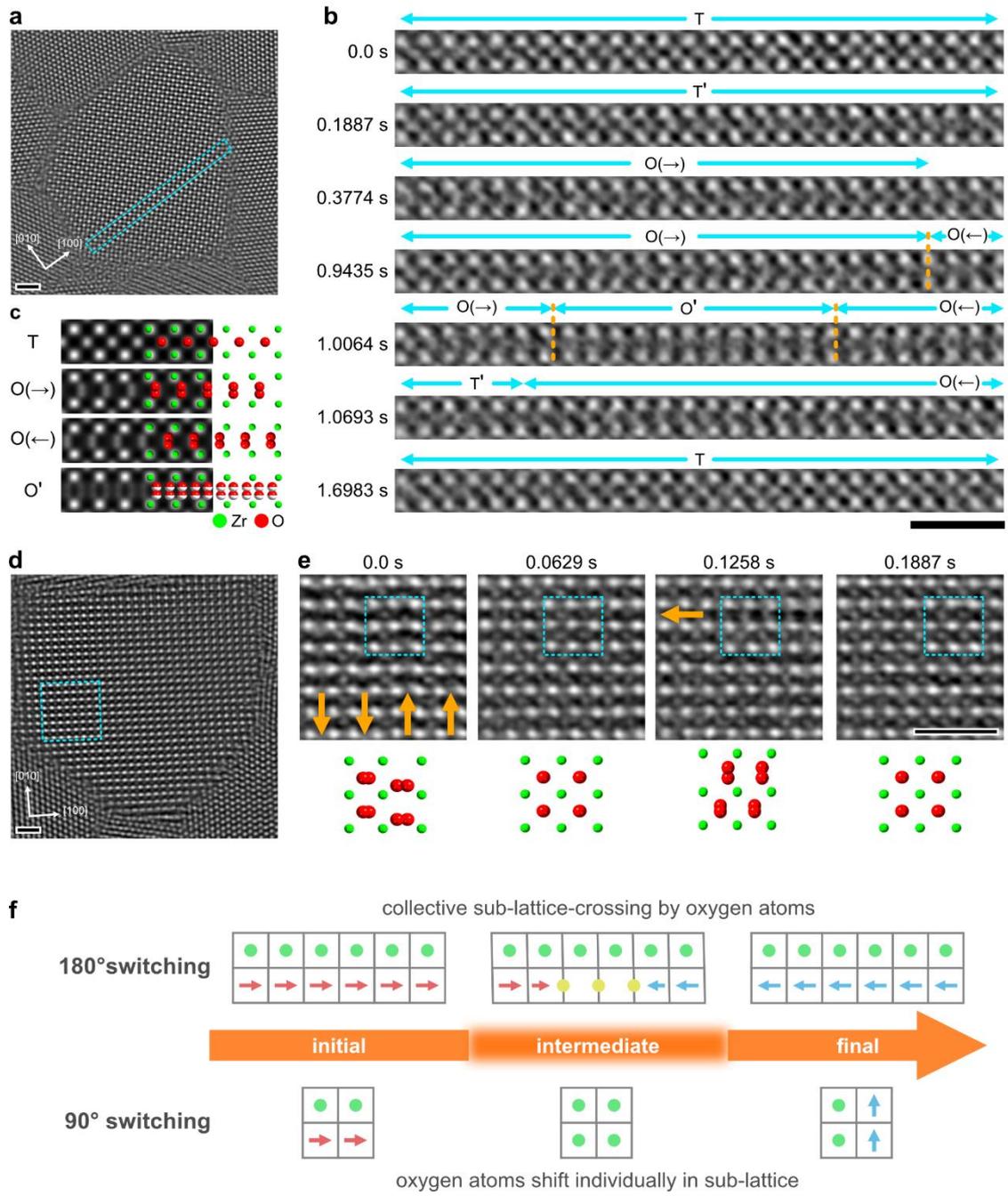

**Figure 3 Observations of intermediate structures during the 180- and 90-degree polarisation switching in ZrO₂.** (a) OBF STEM image of a tetragonal grain, which was obtained by the initial lower probe current illumination (15pA). Scale bar: 1 nm. (b) Time-series of OBF STEM images taken with higher probe current illumination (150 pA), probing the 180-degree polarisation switching with a frame rate of 16 fps or 0.0629 seconds/image. Each frame is obtained by cropping the original dataset from the area highlighted by the cyan dotted box shown in (a). In the series of images, assigned phases are denoted as tetragonal (T), right-polarised orthorhombic (O(→)), left-polarised orthorhombic (O(←)), and nonpolar orthorhombic (O'), respectively. The domain wall between O(→) and O(←) phases is highlighted by the dotted yellow lines. Scale bar: 1 nm. (c) Simulated OBF STEM images of each phase observed in (b) with overlaid structure models, where green and red spheres represent Zr and O sites, respectively. The half-red and half-white spheres indicate the O sites with a half occupancy. (d) OBF STEM image of another tetragonal grain, which is obtained by the initial lower probe current

illumination (15pA). Scale bar: 1 nm. (e) Time-series of OBF STEM images, probing the 90-degree polarisation switching. Frame rate and probe current illumination is the same as (c) and each frame is obtained by cropping from the area highlighted by the cyan dotted box shown in (d). The polarisation directions are indicated by yellow arrows. Schematics of atomic structures are also shown for each frame, corresponding to the area featured by the cyan dotted box shown in (e). Scale bar: 1 nm. (f) Schematic of the structure evolution for 180- and 90-degree switching, which is divided into initial, intermediate, and final structures, showing the different intermediate phases. Grey boxes indicate cation sublattices of the orthorhombic and tetragonal phases and coloured arrows specify the polarised phase and their polarisation direction. Nonpolar phases are denoted with green and yellow dots. The green dots indicate the nonpolar phase with oxygen sites centred in the cation lattice (e.g., tetragonal phase and nonpolar spacer layer in orthorhombic phase), whereas the yellow dots indicate the nonpolar O' phase observed in (b).

**Effect of different cation species**

We observed the polarisation switching pathways in the pure $ZrO_2$ membrane in the previous sections. In addition to the $ZrO_2$, ferroelectric fluorite oxides with other cation species are of significant interest because the other cation species may modify the ferroelectric properties [7,25], with the potential to engineer advanced ferroelectric devices. Here we compare $ZrO_2$ with a HZO membrane, which is a well-known and promising ferroelectric material. In this material, the Hf and Zr are occupying cation sites, which can be considered as a binary of $ZrO_2$ and $HfO_2$, exhibiting robust ferroelectricity with a stable orthorhombic phase [26].

We first obtained an OBF STEM image of a tetragonal grain in the HZO membrane with a lower probe current of 15 pA, which is shown in Fig. 4a. As well as the $ZrO_2$ membrane, the oxygen sites are located at the centre of the cation lattice, showing the nonpolar structure. We then repeated the higher probe current experiments using the same approach and experimental conditions used for the $ZrO_2$ membrane. However, after increasing the probe current to 100 pA, the tetragonal grain immediately changed into the polar orthorhombic phase with just a single scan (denoted as 0 second). Furthermore, unlike the $ZrO_2$ membrane, the orthorhombic phase did not change back to the tetragonal phase, nor was 180-degree polarisation switching observed even after more than 200 seconds of irradiation with 100 pA probe current. At 213 seconds, we observed a 180-degree switching of just a single atomic layer. However, phase transitions back to the tetragonal phase and 90-degree switching were not observed. This result indicates that the reversible nature of the phase transition between nonpolar tetragonal and polar orthorhombic phases is absent in the HZO membrane (at least under this illumination condition). In terms of the polarisation switching, the 180-degree type can still occur but even this is rare compared with the pure $ZrO_2$ membrane, and the 90-degree switching and the phase transition to tetragonal phase did not occur at all.

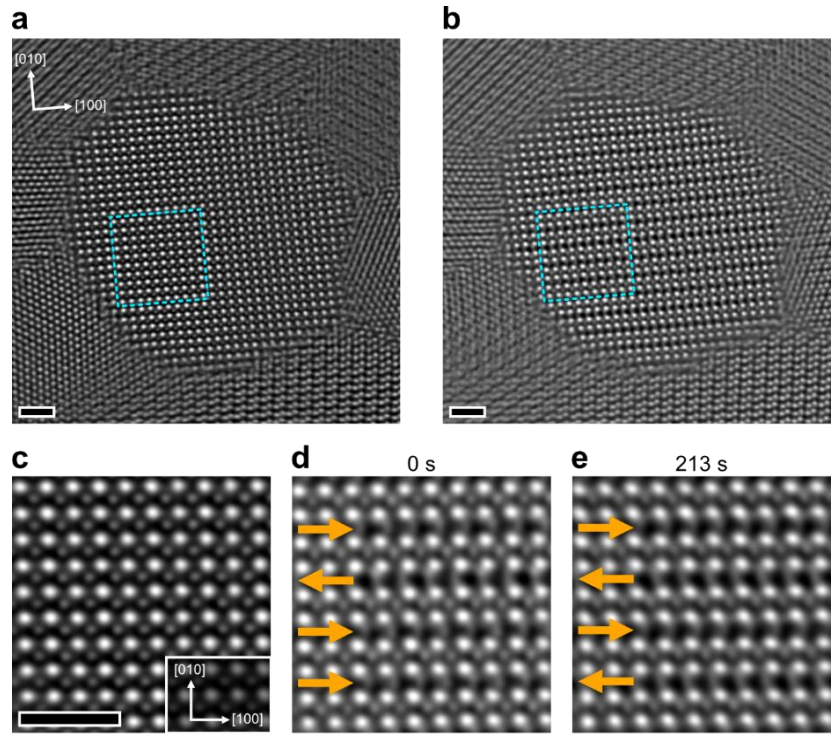

**Figure 4 Observation of non-reversible phase transition and polarisation switching in an HZO membrane.** (a) OBF STEM image of a tetragonal grain in the HZO membrane. (b) OBF STEM image of the polar orthorhombic phase observed in the same grain as (a) just after increasing the probe current into 100 pA. (c, d) Cropped and magnified OBF STEM images obtained from the cyan dotted area in (a) and (b), respectively. Polarisation directions are indicated by yellow arrows. (e) OBF STEM image cropped from the same area as (c, d) after 213 seconds electron illumination with the 100 pA probe current. Scale bars: 1 nm.

# Energetic landscape of structural transitions

## Transition between nonpolar tetragonal and polar orthorhombic phases

To clarify the different structural behaviour of the two membranes, $ZrO_2$ and HZO, under electron irradiation, we performed density functional theory (DFT) calculations. As summarised in Figure 5a and Table 1, we calculated the relative energy of the tetragonal phase with respect to the orthorhombic phase in $ZrO_2$ and $HfO_2$, and the energy barrier to polarisation switching and phase transitions in both materials. It is technically difficult to evaluate the energy landscape of HZO with DFT calculations due to the likely random arrangement of Hf and Zr on the cation site and the periodicity constraints of the calculation. Hence, we calculated energies for pure $ZrO_2$ and $HfO_2$ to obtain insights into HZO, which can be considered as a mixture of $ZrO_2$ and $HfO_2$ [26]. As shown in Fig. 5b, we also schematically summarise the transition/switching pathways accessible in the $ZrO_2$ and $HfO_2$ membranes by indicating which path can be triggered in each material as per the following discussion. As shown in Table 1, the relative energy of the tetragonal phase with respect to the orthorhombic phase of $HfO_2$ is higher than that of $ZrO_2$, indicating that the tetragonal phase is more unstable in $HfO_2$. As shown in Fig. 5a and Table 1, whereas the energy barrier of the phase transition from tetragonal to orthorhombic is comparable between $ZrO_2$ and $HfO_2$, the barrier from orthorhombic to tetragonal shows a much higher value for $HfO_2$ than for $ZrO_2$. These results agree well with the experimental OBF STEM observations: the phase transition between these two phases was observed to be reversible in $ZrO_2$ but irreversible in HZO, that is, in HZO the tetragonal phase can transition to the orthorhombic phase but not *vice versa*. This result also corresponds to our previous findings that the HZO membrane with a 5 nm thickness mostly has orthorhombic grains while the $ZrO_2$ membrane mainly has tetragonal grains, even though both films have approximately the same thickness [9,10].

## 180-degree polarisation switching

For the 180-degree polarisation switching, we observed that the oxygen sites shifted by crossing the cation lattice planes. In the literature describing DFT calculations, two mechanisms are suggested for the 180-degree switching, namely the crossing pathway and the inner pathway [27]. We therefore also calculated the energy diagram of these two paths using the nudged elastic band (NEB) method, which is able to evaluate the ferroelectric transition pathway [28]. In the crossing pathway, the oxygen sites cross between the cation atoms in the (100) and (200) cation lattice planes. Whereas in the inner pathway, the oxygen sites pass through the centre of the cation lattice. From the energy diagram and barrier of the two 180-degree switching mechanisms shown in Fig. 5a, though the minimum energy levels in the middle (50%) of switching paths are comparable for both types, the energy barrier of the crossing pathway is much higher than that of the inner pathway. Our calculations are based on the bulk model, where the calculated structure is periodic and atomic shifting occurs everywhere simultaneously in the bulk. On the other hand, calculations in the literature using the domain wall model, where the large atomic cell is prepared and polarisation switching occurs locally within a single atomic layer, show that the crossing pathway has a lower energy barrier than that for the bulk model [23,27]. It can be said that the energy barrier can be lowered by taking an aperiodic structure locally around the domain wall to initiate the polarisation switching.

In our bulk-model based calculations, we also made a comparison between the two crossing-type pathways with and without a symmetry constraint, as shown in Suppl Figure S5, where the one without symmetry has been employed for Table 1. In the presence of a symmetry constraint, the two oxygen sites, which are visible along the projection direction along [001] in the polarised layer, pass through the cation lattice at the same time, whereas without the symmetry constraint, the oxygen sites pass through one-by-one as seen in the NEB calculation. This indicates that the transient structures during the switching can relax by taking lower-symmetry structures, particularly regarding switching oxygen sites. According to the above discussion, the experimentally observed 180-degree switching shown in Fig. 3b can be interpreted as follows; the oxygen sites passed through the cation lattice one-by-one to lower the energy barrier and this oxygen configuration may be non-uniform along the [001] (film thickness) direction, making the depth-projected oxygen sites look similar to the O' phase with half-occupied oxygen sites as shown in Fig. 3c.

**90-degree polarisation switching**

For the 90-degree polarisation switching in $ZrO_2$, the energy barrier and whole energy diagram show higher values than those of the O→T transition, indicating that the direct 90-degree switching is not energetically preferred and this switching occurs via the O→T transition as observed in the OBF STEM experiment (Fig. 4e), which is also consistent with the STEM observation and DFT calculation in the literature [29]. In the presence of Hf, 180-degree switching has a higher energy barrier than that of $ZrO_2$ as shown in Fig. 5 and Table 1. Moreover, 90-degree switching has an even higher barrier as well as the O→T phase transition, which makes the 90-degree switching less favourable even via the tetragonal phase in $HfO_2$. These DFT results are consistent with the experimental results which show that the HZO membrane has less frequent 180-degree switching than that in $ZrO_2$, and the 90-degree switching was not observed at all.

**Effect of Hf ions in HZO**

Our results indicate that different cation species and/or dopants may substantially affect the switching behaviour and stability of the ferroelectric phases in fluorite oxide materials. In the case of (H)ZO, DFT calculations suggest that the ionic radius of Hf and Zr cations, and the bonding between them and oxygen, may be an important factor: Hf (76 pm) has a smaller ionic radius than Zr (78 pm) [30], resulting in smaller lattice constants of $HfO_2$ compared with $ZrO_2$ [31], which is also confirmed in our DFT calculations (Supplementary Table S1). In addition, it is evaluated that the Hf-O bonding is stronger than that of Zr-O by the difference of their cohesive energy [32], whereby the oxygen sites are less constrained by a softer bonding within a larger cation lattice in $ZrO_2$ compared with $HfO_2$, making it easier for oxygen ions to shift when switching in $ZrO_2$ [33]. In the same way, other species of dopant such as Si, Y, and La at cation sites may also alter the stability of each phase and corresponding switching behaviour as the switching barrier is engineered by the strength of chemical bonds in molecular-type ferroelectrics [34]. Our results highlight the pivotal role of the cation species in tuning the "hardness" of ferroelectricity in fluorite oxides, whereby targeted doping and/or cationic composition may realise a more desirable (e.g., more energy-efficient yet durable) switching response critical for next-generation ferroelectric devices.

**Table 1 DFT-calculated energy barriers of phase transition and polarisation switching in $ZrO_2$ and $HfO_2$.** Relative energy of T phase to O phase, energy barrier of 180-/90-degree polarisation switching and O→T/T→O phase transition in $ZrO_2$ and $HfO_2$ evaluated by NEB method in DFT calculations.

|  | Energy (eV) | |
|---|---|---|
|  | $ZrO_2$ | $HfO_2$ |
| Relative energy of T phase w.r.t. O phase | 0.11 | 0.30 |
| T→O phase transition barrier | 0.14 | 0.19 |
| O→T phase transition barrier | 0.25 | 0.49 |
| 180-degree switching barrier | 0.18 (inner) 0.82 (cross) | 0.34 (inner) |
| 90-degree switching barrier | 0.37 | 0.59 |

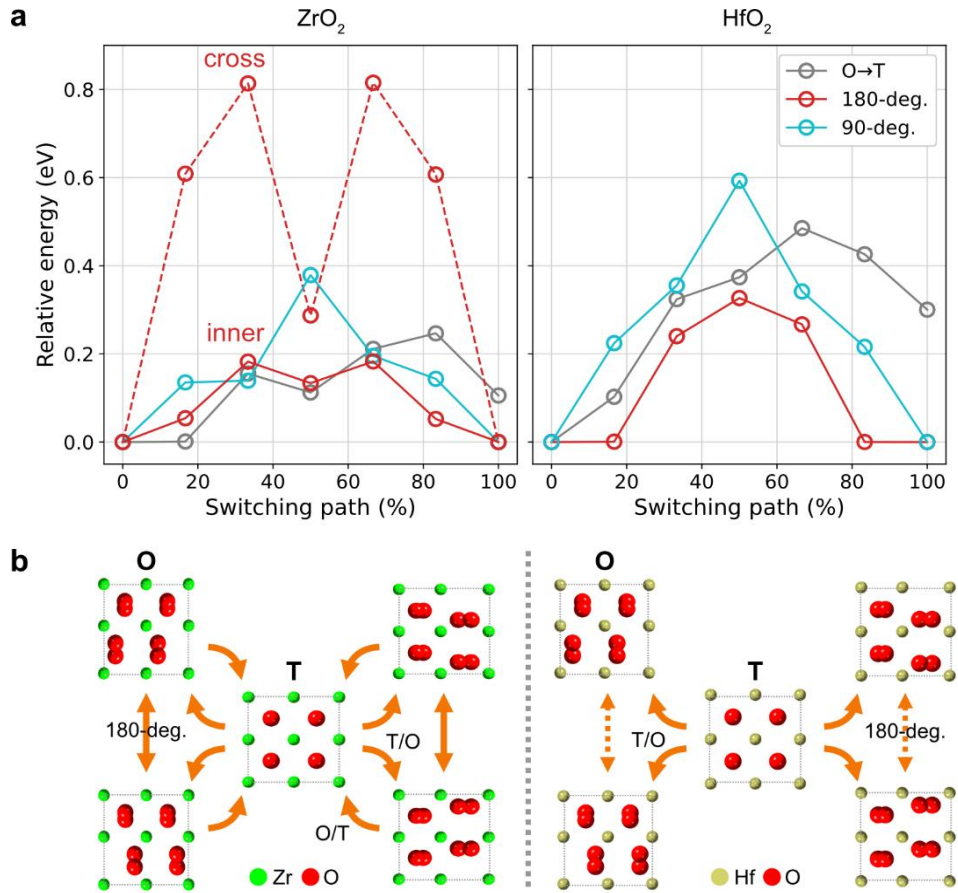

**Figure 5 DFT-calculated energy landscapes of phase transition and polarisation switching in $ZrO_2$ and $HfO_2$.** (a) Energy diagrams of O→T phase transition, 180-degree polarisation switching, and 90-degree polarisation switching in $ZrO_2$ and $HfO_2$. In $ZrO_2$, two types of pathways are calculated for 180-degree switching, namely cross-type and inner-type. (b) Schematic diagram showing the possible transition pathways between nonpolar tetragonal phase and polar orthorhombic phases along four different polarisation directions. The dotted arrows in $HfO_2$ indicate the possible but relatively rare pathway.

# Concluding remarks

Using the OBF STEM technique, we observed directly the evolution of atomic structures during electron illumination in nanometre-thick $ZrO_2$ and $Hf_{0.5}Zr_{0.5}O_2$ (HZO) freestanding membranes, revealing at the atomic level the dynamics of polar-nonpolar phase transitions and polarisation switching. Importantly, we find this to be cation dependent. In the $ZrO_2$ membrane, it was observed that the nonpolar tetragonal phase can be reversibly transformed into the polar orthorhombic phase, where the time-series OBF STEM datasets allowed us to visualise reversible 180- and 90-degree polarisation switching. Furthermore, we clarified the detailed switching pathways by combining higher-speed scanning with OBF STEM: 180-degree polarisation switching occurs via a nonpolar orthorhombic domain wall, while 90-degree switching is facilitated by the nonpolar tetragonal phase, where the dynamic displacement of the oxygen site plays a critical role. Applying the same observation scheme to a $Hf_{0.5}Zr_{0.5}O_2$ (HZO) membrane and complementing the experiments with density functional theory (DFT) calculations, it was revealed that the presence of Hf cation impacts the stability of polar/nonpolar phases and the energy barriers to the phase transition and polarisation switching—rendering the polar orthorhombic phase more energetically favourable in HZO because of the presence of Hf. These findings provide fundamental insights into the atomic scale pathways of polarisation dynamics and phase transition phenomena in ferroelectric fluorite-type materials and the role of the cation, thereby enhancing their potential for advanced ferroelectric device applications.

# Methods

**Fabrication of freestanding (H)ZO membranes**

For the (H)ZO membrane fabrication, we firstly obtained an epitaxially grown (H)ZO/La$_{0.7}$Sr$_{0.3}$MnO$_3$ (LSMO)/SrTiO$_3$ (STO) (001) thin film, at which the (H)ZO and LSMO films are fabricated on an STO (001) substrate by the pulsed laser deposition (PLD) method. The thickness of (H)ZO layer was evaluated by X-ray reflectivity (XRR) measurements. The freestanding (H)ZO membrane was obtained by etching the LSMO layer in 5 wt.% hydrochloric acid (HCl) and solution with 0.05 M potassium iodite (KI). After the etching and exfoliation, the membranes were transferred onto the TEM grids, which have a holey carbon supporting film. We fabricated freestanding five- and ten-nanometre-thick ZrO$_2$ membranes and five-nanometre-thick HZO membrane, respectively, where the detailed fabrication process can be found elsewhere [9,10].

**Atomic resolution structure analysis of (H)ZO membranes**

Atomic resolution STEM datasets were acquired using a Thermo Fisher Scientific Spectra φ FEG TEM equipped with a monochromator and probe and image aberration correctors. All images were collected at an accelerating voltage of 300 kV and probe forming semi-angle of 25 mrad. During the experimental observation, the probe current was adjusted by changing the focus of the monochromator, with all other parameters staying the same. For the OBF STEM imaging, we acquired four quadrant segmented detector datasets using the Panther detector and then reconstructed OBF images via an algorithm that can be referred to in the literature [17]. The ADF images were obtained with an ADF detector simultaneously with the Panther detector datasets. A second order Butterworth low-pass filter [35] was applied to the experimental OBF and ADF images. For the OBF STEM image simulation, firstly the quadrant segmented detector images were calculated by the MuSTEM package [36], being subsequently reconstructed into the OBF image via the same procedure as the experiments. The MuSTEM calculation is based on the multislice algorithm, where the full dynamical scattering effect is incorporated. The crystalline structures used in the image simulation were referred to from the DFT-relaxed crystalline structure files used in our present study (tetragonal and polar orthorhombic phases) and from the literature (antipolar orthorhombic phase) [37].

**DFT calculation**

The DFT calculations were performed to optimise the crystal structures and investigate the energy landscape using the projector augmented-wave (PAW) method implemented in the VASP code. The exchange-correlation term was treated with GGA-PBE. For the PAW potentials, 2s and 2p electrons for O, 4s, 4p, 4d, and 5s electrons for Zr, and 5p, 5d, and 6s electrons for Hf were explicitly treated as valence electrons. The plane-wave cutoff energy was set to 550 eV. Integration in reciprocal space was performed with a grid spacing of 0.3 Å$^{-1}$. Structure optimisation was conducted until all residual forces acting on each atom were less than 0.01 eV/Å. The energy barriers of phase transitions were calculated using the nudged elastic band (NEB) method with five images between initial and final states.

## Acknowledgements

This research was supported under JSPS KAKENHI (Grant No. 22KJ3209 and 23K13553, 21H01810, 23KJ1239, 23H05457, 24H01190, JP23H00241) and grants for the Integrated Research Consortium on Chemical Sciences and the International Collaborative Research Program of the Institute for Chemical Research in Kyoto University from the Ministry of Education, Culture, Sports, Science, and Tech- nology (MEXT) of Japan. The work was also supported by the Japan Science and Technology Agency (JST) as part of PRESTO, Grant No. JPMJPR24H3, and Adopting Sustainable Partnerships for Innovative Research Ecosystem (ASPIRE), Grants No. JPMJAP2312 and No. JPMJAP2314. This work was also supported by The Samco Foundation and by Iketani Science and Technology Foundation. J.E. acknowledges fellowship support from the ARC grant FL220100202. The authors acknowledge the use of instruments and scientific and technical assistance at the Monash Centre for Electron Microscopy (MCEM), Monash University, a node of Microscopy Australia (ROR: 042mm0k03) supported by NCRIS. This research used equipment funded by Australian Research Council (ARC) grants LE0454166 and LE17010118. This work was partially supported by MEXT projects for data-driven materials research (JPMXP1122683430) and "Program for Promoting Researches on the Supercomputer Fugaku" (JPMXP1020230327 and JPMXP1020230325). Computational resources were partially provided by the supercomputer Fugaku at the RIKEN Center for Computational Science (Project IDs: hp230205, hp240212, hp240223).

## Author contributions

K.O. and J.E. conceived, designed, and supervised the project. K.O. performed electron microscopy experiments and simulations, and analysed the data. Y.S. fabricated membrane samples with help from D.K. K.S. performed the density functional theory calculations. K.O. and J.E. prepared the manuscript. All authors contributed to the discussion of the results and revision of the manuscript.

## Competing interests

The authors declare no competing interests.

# References

[1] L.W. Martin, A.M. Rappe, Thin-film ferroelectric materials and their applications, Nat. Rev. Mater. 2 (2016). https://doi.org/10.1038/natrevmats.2016.87.

[2] C.S. Hwang, Prospective of Semiconductor Memory Devices: from Memory System to Materials, Adv. Electron. Mater. 1 (2015) 1–30. https://doi.org/10.1002/aelm.201400056.

[3] M. Dawber, K.M. Rabe, J.F. Scott, Physics of thin-film ferroelectric oxides, Rev. Mod. Phys. 77 (2005) 1083–1130. https://doi.org/10.1103/RevModPhys.77.1083.

[4] J.F. Ihlefeld, D.T. Harris, R. Keech, J.L. Jones, J. Maria, S. Trolier-McKinstry, Scaling Effects in Perovskite Ferroelectrics: Fundamental Limits and Process-Structure-Property Relations, J. Am. Ceram. Soc. 99 (2016) 2537–2557. https://doi.org/10.1111/jace.14387.

[5] U. Schroeder, M.H. Park, T. Mikolajick, C.S. Hwang, The fundamentals and applications of ferroelectric $HfO_2$, Nat. Rev. Mater. 7 (2022) 653–669. https://doi.org/10.1038/s41578-022-00431-2.

[6] M. Bohr, R. Chau, T. Ghani, K. Mistry, The High-k Solution, IEEE Spectr. 44 (2007) 29–35. https://doi.org/10.1109/MSPEC.2007.4337663.

[7] T.S. Böscke, J. Müller, D. Bräuhaus, U. Schröder, U. Böttger, Ferroelectricity in hafnium oxide thin films, Appl. Phys. Lett. 99 (2011) 2–4. https://doi.org/10.1063/1.3634052.

[8] M.H. Park, Y.H. Lee, T. Mikolajick, U. Schroeder, C.S. Hwang, Review and perspective on ferroelectric $HfO_2$-based thin films for memory applications, MRS Commun. 8 (2018) 795–808. https://doi.org/10.1557/mrc.2018.175.

[9] Y. Shen, K. Ooe, X. Yuan, T. Yamada, S. Kobayashi, M. Haruta, D. Kan, Ferroelectric freestanding hafnia membranes with metastable rhombohedral structure down to 1-nm-thick, Nat. Commun. 15 (2024) 1–9. https://doi.org/10.1038/s41467-024-49055-w.

[10] Y. Shen, K. Ooe, K. Shitara, S. Kobayashi, T. Yoshimura, T. Yamada, L. Xie, Y. Shimakawa, D. Kan, Ultrathin freestanding membranes of $ZrO_2$ with metastable structures and strain-dependent electrical properties, Phys. Rev. Mater. 9 (2025) 024411. https://doi.org/10.1103/PhysRevMaterials.9.024411.

[11] J. (Jimmy) Liu, Advances and Applications of Atomic-Resolution Scanning Transmission Electron Microscopy, Microsc. Microanal. 27 (2021) 943–995. https://doi.org/10.1017/S1431927621012125.

[12] S. Kobayashi, K. Inoue, T. Kato, Y. Ikuhara, T. Yamamoto, Multiphase nanodomains in a strained $BaTiO_3$ film on a $GdScO_3$ substrate, J. Appl. Phys. 123 (2018). https://doi.org/10.1063/1.5012545.

[13] H. Zhong, M. Li, Q. Zhang, L. Yang, R. He, F. Liu, Z. Liu, G. Li, Q. Sun, D. Xie, F. Meng, Q. Li, M. He, E. Guo, C. Wang, Z. Zhong, X. Wang, L. Gu, G. Yang, K. Jin, P. Gao, C. Ge, Large-Scale $Hf_{0.5}Zr_{0.5}O_2$ Membranes with Robust Ferroelectricity, Adv. Mater. 34 (2022) 2109889. https://doi.org/10.1002/adma.202109889.

[14] X. Li, H. Zhong, T. Lin, F. Meng, A. Gao, Z. Liu, D. Su, K. Jin, C. Ge, Q. Zhang, L. Gu, Polarization Switching and Correlated Phase Transitions in Fluorite-Structure $ZrO_2$ Nanocrystals, Adv. Mater. 35 (2023) 1–9. https://doi.org/10.1002/adma.202207736.

[15] X. Li, Z. Liu, A. Gao, Q. Zhang, H. Zhong, F. Meng, T. Lin, S. Wang, D. Su, K. Jin, C. Ge, L. Gu, Ferroelastically protected reversible orthorhombic to monoclinic-like phase transition in $ZrO_2$ nanocrystals, Nat. Mater. 23 (2024) 1077–1084. https://doi.org/10.1038/s41563-024-01853-9.

[16] Y. Cheng, Z. Gao, K.H. Ye, H.W. Park, Y. Zheng, Y. Zheng, J. Gao, M.H. Park, J.-H. Choi, K.-H. Xue, C.S. Hwang, H. Lyu, Reversible transition between the polar and antipolar phases and its implications for wake-up and fatigue in $HfO_2$-based ferroelectric thin film, Nat. Commun. 13 (2022) 645. https://doi.org/10.1038/s41467-022-28236-5.

[17] K. Ooe, T. Seki, Y. Ikuhara, N. Shibata, Ultra-high contrast STEM imaging for segmented/pixelated detectors by maximizing the signal-to-noise ratio, Ultramicroscopy. 220 (2021) 113133. https://doi.org/10.1016/j.ultramic.2020.113133.

[18] H.J. Lee, M. Lee, K. Lee, J. Jo, H. Yang, Y. Kim, S.C. Chae, U. Waghmare, J.H. Lee, Scale-free ferroelectricity induced by flat phonon bands in $HfO_2$, Science. 369 (2020) 1343–1347. https://doi.org/10.1126/SCIENCE.ABA0067.

[19] S. Anada, T. Nagase, H. Yasuda, H. Mori, Electron-irradiation-induced phase transition in $Cr_2M$ (M = Ti and Al) intermetallic compounds, J. Alloys Compd. 579 (2013) 646–653. https://doi.org/10.1016/j.jallcom.2013.06.122.

[20] C.W. Huang, S.S. Kuo, C.L. Hsin, Electron-beam-induced phase transition in the transmission electron microscope: The case of $VO_2(B)$, CrystEngComm. 20 (2018) 6857–6860. https://doi.org/10.1039/c8ce01536h.

[21] J. Wei, B. Feng, R. Ishikawa, T. Yokoi, K. Matsunaga, N. Shibata, Y. Ikuhara, Direct imaging of atomistic grain boundary migration, Nat. Mater. 20 (2021) 951–955. https://doi.org/10.1038/s41563-020-00879-z.

[22] R.-J. Jiang, Y. Cao, W.-R. Geng, M.-X. Zhu, Y.-L. Tang, Y.-L. Zhu, Y. Wang, F. Gong, S.-Z. Liu, Y.-T. Chen, J. Liu, N. Liu, J.-H. Wang, X.-D. Lv, S.-J. Chen, X.-L. Ma, Atomic Insight into the Successive Antiferroelectric–Ferroelectric Phase Transition in Antiferroelectric Oxides, Nano Lett. 23 (2023) 1522–1529. https://doi.org/10.1021/acs.nanolett.2c04972.

[23] M. Noor, M. Bergschneider, J. Kim, N. Afroze, A.I. Khan, S.C. Chang, U.E. Avci, A.C. Kummel, K. Cho, Nearly Barrierless Polarization Switching Mechanisms in $ZrO_2$ Having Perpendicular In-Plane Domain Walls, ACS Appl. Mater. Interfaces. (2024). https://doi.org/10.1021/acsami.4c07728.

[24] B. N.A, P. S.V., V. L.F., New high pressure modifications of $ZrO_2$ and $HfO_2$, Geochemistry Int. 4 (1967) 557.

[25] J. Müller, U. Schröder, T.S. Böscke, I. Müller, U. Böttger, L. Wilde, J. Sundqvist, M. Lemberger, P. Kücher, T. Mikolajick, L. Frey, Ferroelectricity in yttrium-doped hafnium oxide, J. Appl. Phys. 110 (2011) 1–5. https://doi.org/10.1063/1.3667205.

[26] J. Müller, T.S. Böscke, U. Schröder, S. Mueller, D. Bräuhaus, U. Böttger, L. Frey, T. Mikolajick, Ferroelectricity in simple binary $ZrO_2$ and $HfO_2$, Nano Lett. 12 (2012) 4318–4323. https://doi.org/10.1021/nl302049k.

[27] Y. Wu, Y. Zhang, J. Jiang, L. Jiang, M. Tang, Y. Zhou, M. Liao, Q. Yang, E.Y. Tsymbal, Unconventional Polarization-Switching Mechanism in (Hf, Zr)$O_2$ Ferroelectrics and Its Implications, Phys. Rev. Lett. 131 (2023) 226802. https://doi.org/10.1103/PhysRevLett.131.226802.

[28] X.Y. Li, Q. Yang, J.X. Cao, L.Z. Sun, Q.X. Peng, Y.C. Zhou, R.X. Zhang, Domain Wall Motion in Perovskite Ferroelectrics Studied by the Nudged Elastic Band Method, J. Phys. Chem. C. 122 (2018) 3091–3100. https://doi.org/10.1021/acs.jpcc.7b11330.

[29] S. Wang, X. Li, Z. Liu, A. Gao, Q. Zhang, T. Lin, H. Zhong, D. Su, K. Jin, C. Ge, L. Gu, Unconventional Ferroelectric-Ferroelastic Switching Mediated by Non-Polar Phase in Fluorite Oxides, Adv. Mater. 2415131 (2025) 1–9. https://doi.org/10.1002/adma.202415131.

[30] R.D. Shannon, Revised effective ionic radii and systematic studies of interatomic distances in halides and chalcogenides, Acta Crystallogr. Sect. A. 32 (1976) 751–767. https://doi.org/10.1107/S0567739476001551.

[31] S.E. Reyes-Lillo, K.F. Garrity, K.M. Rabe, Antiferroelectricity in thin-film $ZrO_2$ from first principles, Phys. Rev. B. 90 (2014) 140103. https://doi.org/10.1103/PhysRevB.90.140103.

[32] N.I. Medvedeva, V.P. Zhukov, M.Y. Khodos, V.A. Gubanov, The Electronic Structure and Cohesive Energy of $HfO_2$, $ZrO_2$, $TiO_2$, and $SnO_2$ Crystals, Phys. Status Solidi. 160 (1990) 517–527. https://doi.org/10.1002/pssb.2221600213.

[33] X. Dou, W. Wei, P. Sang, L. Tai, X. Li, X. Zhan, J. Wu, J. Chen, Polarization switching pathways of ferroelectric Zr-doped $HfO_2$ based on the first-principles calculation, Appl. Phys. Lett. 124 (2024). https://doi.org/10.1063/5.0194409.

[34] Y. Hu, K. Parida, H. Zhang, X. Wang, Y. Li, X. Zhou, S.A. Morris, W.H. Liew, H. Wang, T. Li, F. Jiang, M. Yang, M. Alexe, Z. Du, C.L. Gan, K. Yao, B. Xu, P.S. Lee, H.J. Fan, Bond engineering of molecular ferroelectrics renders soft and high-performance piezoelectric energy harvesting materials, Nat. Commun. 13 (2022) 1–10. https://doi.org/10.1038/s41467-022-33325-6.

[35] S. Butterworth, On the Theory of Filter Amplifiers, Wirel. Eng. 7 (1930) 536–541.

[36] L.J. Allen, A.J. D'Alfonso, S.D. Findlay, Modelling the inelastic scattering of fast electrons, Ultramicroscopy. 151 (2015) 11–22. https://doi.org/10.1016/j.ultramic.2014.10.011.

[37] A. Kersch, M. Falkowski, New Low-Energy Crystal Structures in $ZrO_2$ and $HfO_2$, Phys. Status Solidi - Rapid Res. Lett. 15 (2021) 10–13. https://doi.org/10.1002/pssr.202100074.

Supplementary materials for

# Direct observation of cation-dependent polarisation switching dynamics in fluorite ferroelectrics

Kousuke Ooe[1,2*], Yufan Shen[3], Kazuki Shitara[2], Shunsuke Kobayashi[2], Yuichi Shimakawa[3], Daisuke Kan[3†], Joanne Etheridge[1,4]

[1]School of Physics and Astronomy, Monash University, Clayton, VIC 3800, Australia.
[2]Nanostructures Research Laboratory, Japan Fine Ceramics Center, Nagoya, Aichi 456-8587, Japan.
[3]Institute for Chemical Research, Kyoto University, Uji, Kyoto 611-0011, Japan.
[4]Monash Centre for Electron Microscopy, Monash University, Clayton, VIC 3800, Australia.
* Corresponding author. Email: kousuke.ooe@monash.edu (K.O.)
† Present address: Division of Applied Chemistry, Graduate School of Engineering, The University of Osaka, Suita, Osaka 565-0871, Japan.

**The PDF file includes:**

    Supplementary notes 1 to 3
    Supplementary figures S1 to S5
    Supplementary table S1
    Supplementary movies S1 to S6
    References in Supplementary Materials

**Supplementary Note 1. OBF STEM image simulation of ZrO2 sample**

To ensure we correctly interpret the atomic structures, including oxygen sites, from the OBF STEM images, we performed a comprehensive multislice calculation to generate the scattered electron wavefield and then used this to generate an OBF STEM image of tetragonal, polar orthorhombic, and antipolar orthorhombic phases in $ZrO_2$, to compare with the experimental data shown in Fig. 1. We first simulated data assuming a quadrant segmented detector and observation of 5-nm thick tetragonal, 5-nm thick polar orthorhombic, and 10-nm thick antipolar orthorhombic phases with a 300 kV accelerating voltage and 25 mrad probe forming aperture. The OBF STEM images were then reconstructed by using the simulated segmented detector dataset. Figure S1 shows the simulated OBF STEM images corresponding to each phase and thickness, where the atomic structure including both Zr and O sites is clearly visualised. Particularly focusing on the oxygen sites, we can distinguish the location of them with reference to Zr cation lattice: oxygen is located at the centre of the cation lattice in the tetragonal phase, and off-centred in the polarised layers of (anti)polar orthorhombic phase. The polarisation direction can also be identified by which direction the oxygen sites are shifted to.

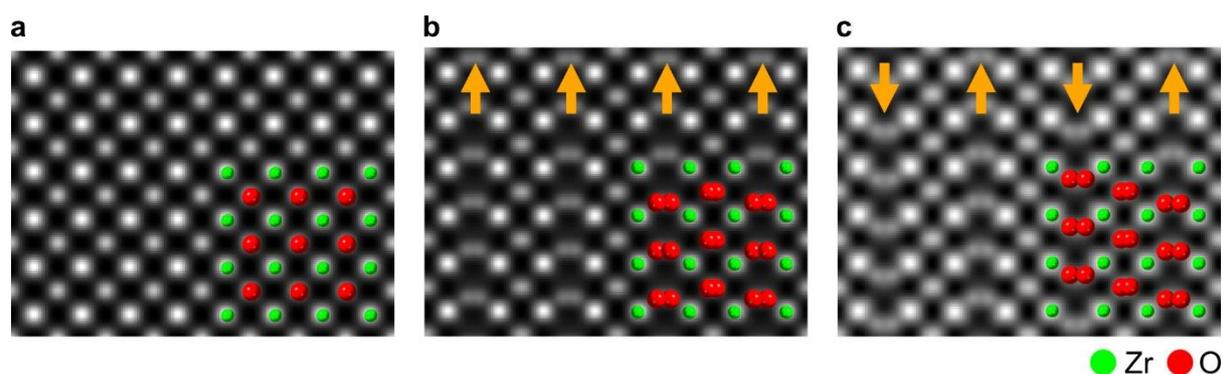

Figure S1 Simulated OBF STEM images of (a) tetragonal phase, (b) polar orthorhombic, and (c) antipolar orthorhombic phases in $ZrO_2$. In accordance with the experimental images shown in Fig. 1, sample thickness for the simulation is 5 nm for (a) and (b), and 10 nm for (c). The Zr and O sites are indicated by green and red spheres, and polarisation direction is shown in yellow arrows.

# Supplementary Note 2. Cation interatomic distance measurements from ADF STEM images

As shown in our present study, OBF STEM can visualise oxygen sites in (H)ZO membranes, which is powerful for investigating the phase transition and polarisation switching at the atomic scale. While the evolution of the phase transition and switching is manifested by the distinct oxygen shifts, these structural changes are also accompanied by a change in the cation distribution. As the ADF STEM image, which excels in imaging heavy atomic number sites, was also recorded simultaneously with the segmented detector dataset, we can analyse the cation interatomic distances from the ADF images and correlate this with the oxygen positions identified in the OBF-STEM images.

The simultaneously acquired ADF and OBF images are shown in Figure S2. This data is the same as Fig. 2a, where the tetragonal phase was observed with a lower probe current. In the ADF image shown in Fig. S2a and S2b, the Zr sites are visualised, but the O sites are absent in the image contrast. While the OBF image shown in Fig. S2c and S2d shows the atomic structure, including both Zr and O sites, the intensity maxima at the Zr sites appear to have a "slight" four-fold symmetry. The segmented-detector used for the data collection has a quadrant-shaped detection channel, and it is known that the obtainable OBF image contrast may have an image artifact that reflects the symmetry of the detector used [1]. While this artifact is negligible for the light element sites such as oxygen, it becomes noticeable for the heavy element sites and/or for larger sample thicknesses. The (H)ZO membrane studied here has 5-10 nm thickness, which is in the range of validity of the OBF-STEM approximations and does not affect the interpretation of the oxygen sites. In addition, in bright-field-signal based imaging techniques (e.g., OBF STEM and ABF STEM etc.), if the incident beam direction is not well-aligned with the zone-axis of the crystal, accurate measurement of atomic positions and interatomic distances may be hindered [2]. As the freestanding membrane (H)ZO is a nano-sized polycrystalline sample and precise tilt-alignment is technically challenging, ADF imaging is more suitable for the detailed and quantitative measurement of cation sites position and distance.

We measured the cation interatomic distances along [100] and [010] directions from time-series ADF images, as shown in Figure S3. For measuring the cation interatomic distances, we first identify the peak positions in the ADF STEM intensity distribution using two-dimensional Gaussian fitting via Atomap software [3], where the distances between the fitted positions along [100] and [010] directions are measured. These time-series datasets are acquired simultaneously with the OBF images shown in Fig. 2e, where the 90-degree polarisation switching was observed. At 216 seconds, as the OBF image showed, the observed grain exhibits a tetragonal structure. The cation interatomic distance measured by the ADF image shows uniform values, which is also visualised by the histogram of distances. We also fit two-dimensional Gaussian functions to the histograms along each direction, showing almost the same peak positions, which corresponds to the uniform cation lattice structure. At 219 seconds, the atomic structure changed to orthorhombic phase polarised along the [100] direction, where the cation interatomic distances show a homogeneous value along the [100] direction but two distinct values along the [010] direction. As discussed in Fig. 1a, the orthorhombic phase has uniform cation lattice distances parallel to the polarisation direction and two alternating lattice distances (wide and narrow layers) perpendicular to it. These two distances are also evident in the corresponding histogram and fitted Gaussian functions in Fig S3. At 222 seconds, the polarisation direction is changed to the [010] direction and the two alternating interatomic distances are confirmed in the [100] direction. After an extra scan, at 225 seconds, the atomic structure has changed back to the tetragonal structure observed at 216 seconds, as confirmed by the uniform cation interatomic distances along both [100] and [010] directions.

Furthermore, at 231 seconds, the grain had both tetragonal and orthorhombic structures polarised along [100] direction within the same grain, which is manifested by the interatomic distances and their histogram as shown in Figure S4a and S4b. We magnify the interface region between the two phases, which is featured by the red box in Fig. S4a, as shown in Fig. S4c. In the measured cation interatomic distance along [010] direction in this region, the cation lattice has two alternating distances in the bottom side, while it has uniform distances in the upper side, showing the existence of the interface. The OBF image in the same field of view is shown in Fig. S4d, which clearly visualises the two different phases by the location of oxygen sites. These two phases are connected to each other coherently, implying a low interface energy.

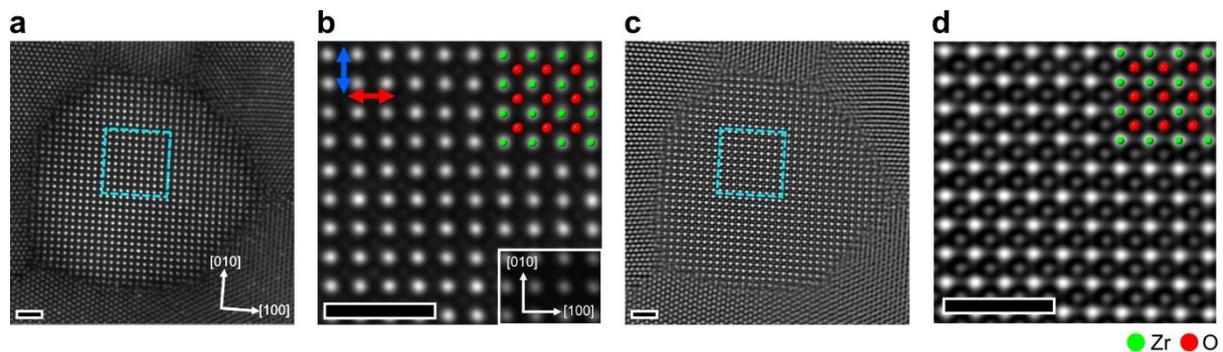

Figure S2 Comparison between (a, b) ADF- and (c, d) OBF-STEM images obtained simultaneously from a tetragonal (T) phase grain in the $ZrO_2$ membrane. (b) and (d) are enlarged views of the cyan square region in (a) and (c), respectively. Atomic structure models are overlayed with red dots referring to the O sites and green dots to the Zr sites. Scale bar: 1 nm.

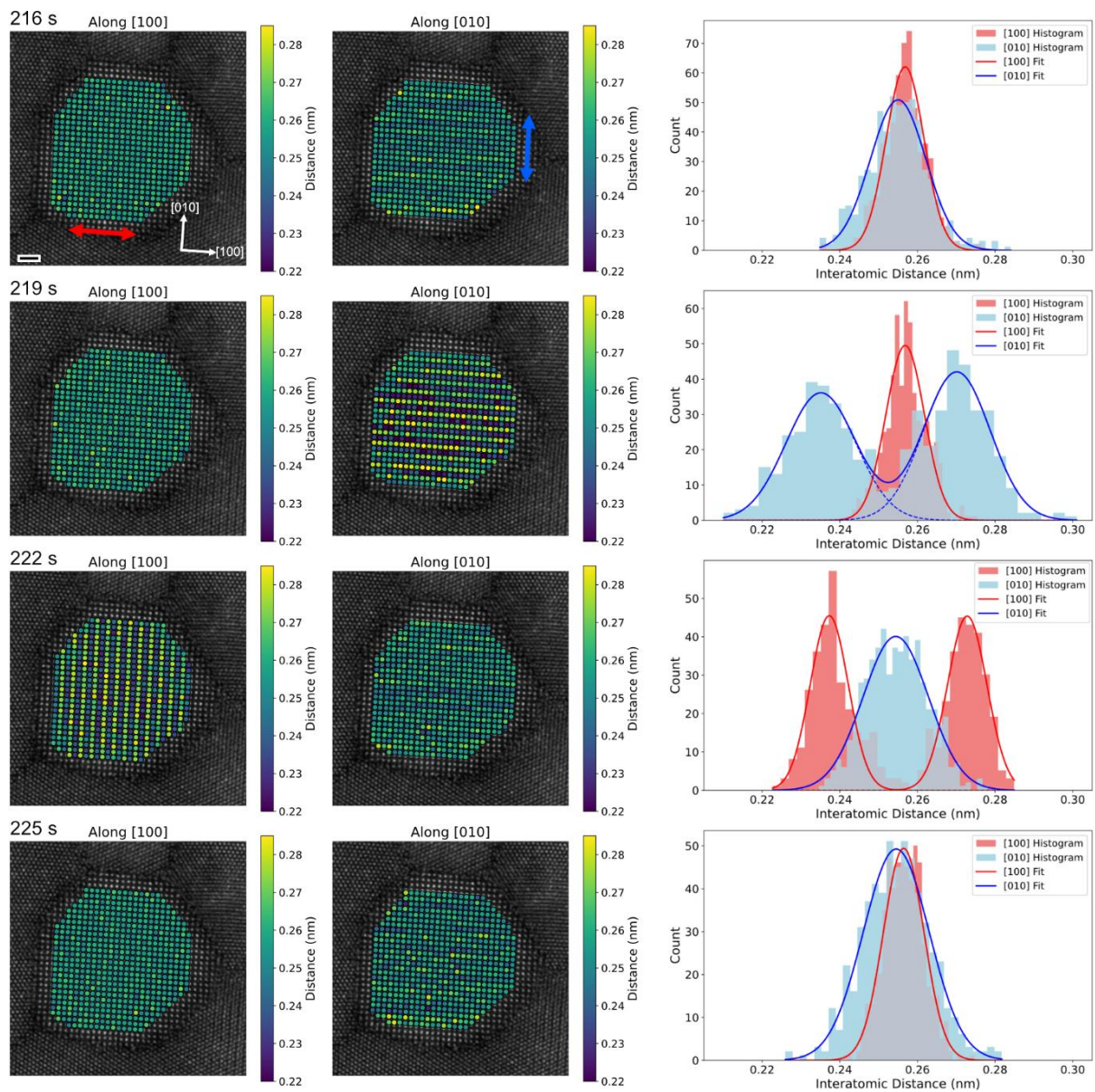

Figure S3 Zr cation interatomic distances measured from time-series ADF-STEM images of the $ZrO_2$ membrane. The distances are measured along the [100] and [010] directions, and are indicated by the coloured dots in between the cation sites. At each time, we calculated histograms of measured distances along both directions (red and blue colours for [100] and [010] directions, respectively). Fitted Gaussian functions are also plotted for each histogram. Scale bar: 1 nm.

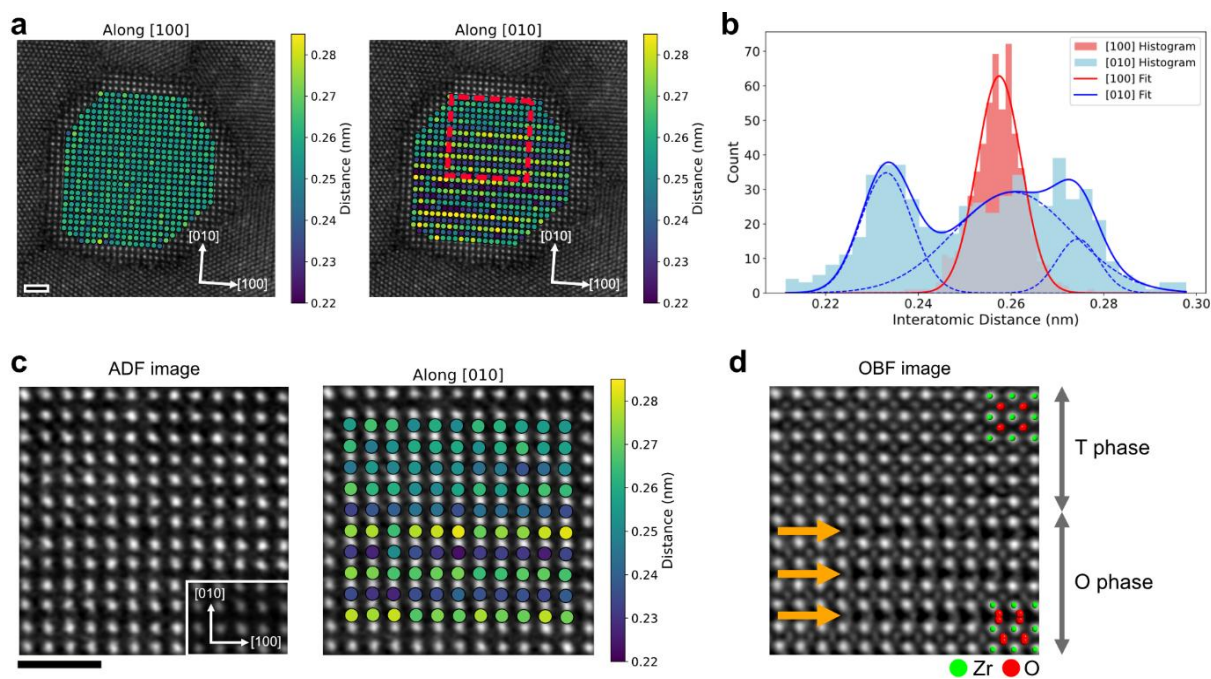

Figure S4 Cation interatomic distance measurement of the structure with T and O phases coexisting within a single grain (as observed after 231 seconds in the series in Figures S3). (a) Interatomic distance map along the [100] and [010] directions. The red dotted box indicates the area featured in (c). Scale bar: 1 nm. (b) Histogram of measured interatomic distances along [100] (red) and [010] (blue) directions. Gaussian functions fitted to the histogram are plotted for each direction. (c) The ADF image cropped and magnified from the red dotted box area shown in (a). The cation interatomic distances along the [010] direction is also shown. Scale bar: 1 nm. (d) The OBF image obtained simultaneously with (c). The polarisation direction within the polar orthorhombic phase is indicated by yellow arrows, and Zr and O sites are indicated by green and red dots, respectively.

# Supplementary Note 3. Intermediate structures in the energy diagram obtained from DFT calculations

To understand the structure evolution during the phase transition and polarisation switching observed experimentally in the (H)ZO membranes, we performed DFT calculations using the nudged elastic band (NEB) method, which can calculate the intermediate structures within an atomic structure evolution including polarisation switching. Before undertaking the NEB calculation, we first evaluated the relaxation of the unit-cell parameters using the DFT calculation, which is summarised in Table S1. As discussed in the literature [4], the $ZrO_2$ shows larger lattice constant than that of $HfO_2$ (HO) in both tetragonal and orthorhombic phases due to the smaller $Hf^{4+}$ ionic radius.

The structure evolution during the O→T phase transition and 180-/90-deg. polarisation switching was then calculated based on the NEB method as shown in Figure S5. For the NEB calculation of each structure evolution, the intermediate structures are explored by sampling five points between the given initial and final structures, resulting in seven structures per one pathway. For the 180-degree switching, we performed the calculation for inner- and cross-type switching, which has been suggested as two possible pathways for 180-degree switching in the DFT literature [5,6]. In the inner-type, the polarised oxygen sites are shifting inside the cation lattice, whereas in the cross-type, these oxygen sites are shifting by crossing the cation lattice plane. In our OBF STEM experiment shown in Fig. 3b, the cross-type pathway was observed. So for the cross-type pathway, we also performed the calculation without a symmetry-constraint. In the cross-type pathway with the symmetry constraint, the two oxygen sites are shifting and crossing the cation lattice simultaneously. Once the symmetry constraint is removed, however, the two oxygen sites are crossing the cation plane one-by-one rather than crossing at the same time. Comparing these cross-type structure evolutions with/without the symmetry constraint, it can be said that the energy barrier can relax by lowering the structure symmetry regarding switching oxygen sites, which may lead to the half occupancy of oxygen sites in the nonpolar (*Pbcm*) orthorhombic phase. This is consistent with the lowered OBF image intensity of oxygen sites as discussed in Fig. 3b.

Table S1 DFT-calculated lattice constants of orthorhombic and tetragonal phases in pure $ZrO_2$ and $HfO_2$. The tetragonal lattice here is defined so that it corresponds to the orthorhombic phase.

|  | $ZrO_2$ ortho. | $ZrO_2$ tetra. | $HfO_2$ ortho. | $HfO_2$ tetra. |
| --- | --- | --- | --- | --- |
| a (Å) | 5.0524 | 5.0729 | 4.998 | 5.0231 |
| b (Å) | 5.0712 | 5.0729 | 5.0211 | 5.0231 |
| c (Å) | 5.2582 | 5.1775 | 5.2045 | 5.133 |
| alpha (deg.) | 90 | 90 | 90 | 90 |
| beta (deg.) | 90 | 90 | 90 | 90 |
| gamma (deg.) | 90 | 90 | 90 | 90 |

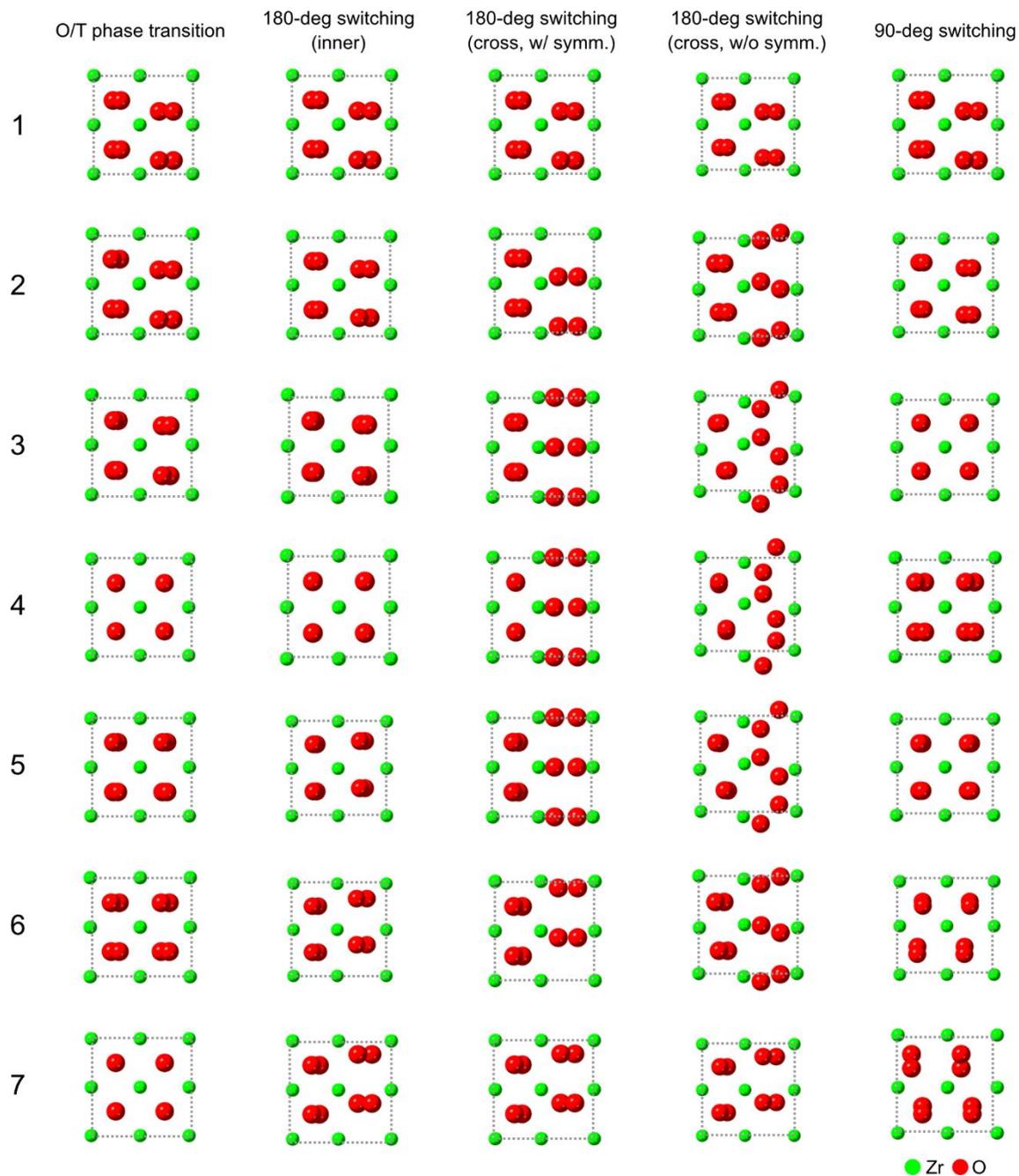

Figure S5 A series of DFT-calculated structure evolutions during O→T phase transition, 180-degree polarisation switching (inner-type, cross-type with symmetry, and cross-type without symmetry), and 90-degree polarisation switching. The DFT calculation assumes the bulk model, where the atomic structure is periodic, indicating the atomic displacement according to the evolution occurs simultaneously everywhere in the bulk. All atomic structures are shown as a projection along [001] direction.

Movie S1. Time-series OBF STEM images of a tetragonal grain of the $ZrO_2$ membrane with a lower electron dose illumination.

Movie S2. Time-series OBF STEM images of the tetragonal grain of the $ZrO_2$ membrane with a higher electron dose illumination (0-87 seconds, 0.33 frame per second). The tetragonal grain starts to change to the polar orthorhombic phase reversibly.

Movie S3. Time-series OBF STEM images of the tetragonal grain of the $ZrO_2$ membrane with a higher electron dose illumination (90-177 seconds, 0.33 frame per second). Reversible phase transition between tetragonal and orthorhombic phases and 180-degree polarisation switching are present.

Movie S4: Time-series OBF STEM images of the originally tetragonal grain of the $ZrO_2$ membrane with a higher electron dose illumination (180-267 seconds, 0.33 frame per second). In addition to the phase transition and 180-degree polarisation switching, 90-degree polarisation switching also appears.

Movie S5: Higher-speed time-series OBF STEM images of a tetragonal grain of the $ZrO_2$ membrane with a higher electron dose illumination (0-2.4531 seconds, 16 frame per second). The movie shows the cropped area as shown in Fig. 3b, visualising the 180-degree polarisation switching.

Movie S6: Higher-speed time-series OBF STEM images of an originally tetragonal grain of the $ZrO_2$ membrane with a higher electron dose illumination (0-2.4531 seconds, 16 frame per second). The movie shows the cropped area as shown in Fig. 3e, visualising the 90-degree polarisation switching.

# References

[1] K. Ooe, T. Seki, Y. Ikuhara, N. Shibata, Ultra-high contrast STEM imaging for segmented/pixelated detectors by maximizing the signal-to-noise ratio, Ultramicroscopy 220 (2021) 113133. https://doi.org/10.1016/j.ultramic.2020.113133.

[2] S. Kobayashi, A. Kuwabara, C.A.J. Fisher, Y. Ikuhara, Atomic-Scale Analysis of Biphasic Boundaries in the Lithium-Ion Battery Cathode Material LiFePO$_4$, ACS Appl. Energy Mater. 3 (2020) 8009–8016. https://doi.org/10.1021/acsaem.0c01408.

[3] M. Nord, P.E. Vullum, I. MacLaren, T. Tybell, R. Holmestad, Atomap: a new software tool for the automated analysis of atomic resolution images using two-dimensional Gaussian fitting, Adv. Struct. Chem. Imaging 3 (2017). https://doi.org/10.1186/s40679-017-0042-5.

[4] S.E. Reyes-Lillo, K.F. Garrity, K.M. Rabe, Antiferroelectricity in thin-film ZrO$_2$ from first principles, Phys. Rev. B 90 (2014) 140103. https://doi.org/10.1103/PhysRevB.90.140103.

[5] Y. Wu, Y. Zhang, J. Jiang, L. Jiang, M. Tang, Y. Zhou, M. Liao, Q. Yang, E.Y. Tsymbal, Unconventional Polarization-Switching Mechanism in (Hf, Zr)O$_2$ Ferroelectrics and Its Implications, Phys. Rev. Lett. 131 (2023) 226802. https://doi.org/10.1103/PhysRevLett.131.226802.

[6] X. Dou, W. Wei, P. Sang, L. Tai, X. Li, X. Zhan, J. Wu, J. Chen, Polarization switching pathways of ferroelectric Zr-doped HfO$_2$ based on the first-principles calculation, Appl. Phys. Lett. 124 (2024). https://doi.org/10.1063/5.0194409.